\documentstyle[12pt,aaspp4,psfig]{article}

\def\gro17{GRO~J1719$-$24}
\def\ea{\hbox{et~al.}}
\lefthead{F. van der Hooft \ea}
\righthead{Hard X-ray lags in GRO~J1719$-$24}

\begin{document}

\title{Hard X-ray lags in GRO~J1719$-$24} 

\author{
F.~van~der~Hooft\altaffilmark{1},
C.~Kouveliotou\altaffilmark{2,3},
J.~van~Paradijs\altaffilmark{1,4},
W.S.~Paciesas\altaffilmark{4},
W.H.G.~Lewin \altaffilmark{5},
M.~van~der~Klis\altaffilmark{1,6},
D.J.~Crary\altaffilmark{3,7},
M.H.~Finger\altaffilmark{2,3},
B.A.~Harmon \altaffilmark{3},
and S.N.~Zhang\altaffilmark{2,3}
}

\altaffiltext{1}{Astronomical Institute ``Anton Pannekoek'', University of Amsterdam
and Center for High Energy Astrophysics, Kruislaan 403, NL-1098 SJ Amsterdam, The Netherlands}
\altaffiltext{2}{Universities Space Research Association, Huntsville, AL 35806, USA} 
\altaffiltext{3}{ES-84, NASA/Marshall Space Flight Center, Huntsville, AL 35812, USA} 
\altaffiltext{4}{Department of Physics, University of Alabama in Huntsville, Huntsville, AL 35899, USA} 
\altaffiltext{5}{Massachusetts Institute of Technology, 37-627 Cambridge, MA 02139, USA}
\altaffiltext{6}{Department of Astronomy, University of California, Berkeley, Berkeley, CA 94720, USA}
\altaffiltext{7}{NAS/NRC Research Associate, NASA Code ES-84, Marshall Space Flight Center, Huntsville, AL 35812, USA}

%
%
\begin{abstract} 
We have used the Fourier cross spectra of \gro17, as obtained with BATSE, to estimate the phase 
lags between the X-ray flux variations in the 20--50 and 50--100 keV 
energy bands as a function of Fourier frequency in the interval 0.002--0.488 Hz. 
Our analysis covers the entire $\sim$\,80 day X-ray outburst of this black-hole candidate, 
following the first X-ray detection on 1993 September 25. 
The X-ray variations in the 50--100 keV band lag those in the 20--50 keV energy band by an 
approximately constant phase difference of 0.072\,$\pm$\,0.010 rad in the frequency interval 0.02--0.20 Hz. 
The peak phase lags in the interval 0.02--0.20 Hz are about twice those of Cyg~X-1 and GRO~J0422$+$32.
These results are consistent with 
models for Comptonization regions composed of extended non-uniform clouds around the central source. 
\end{abstract} 


%
%

\section{Introduction} 

\begin{figure*} 
\hbox{
\psfig{figure=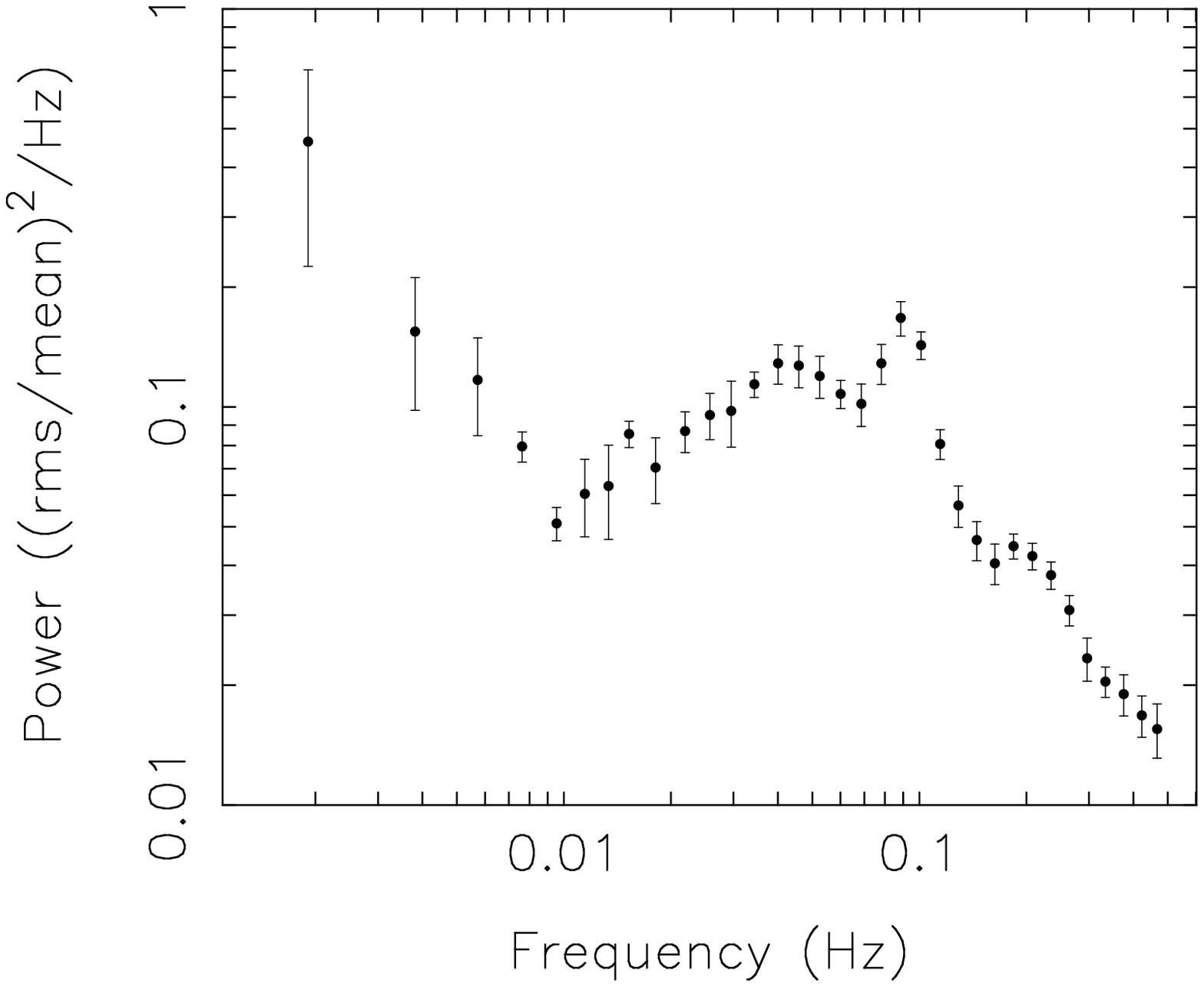,width=8.0cm,bbllx=179pt,bblly=60pt,bburx=565pt,bbury=529pt,angle=-90} 
\hspace{0.1cm}
\psfig{figure=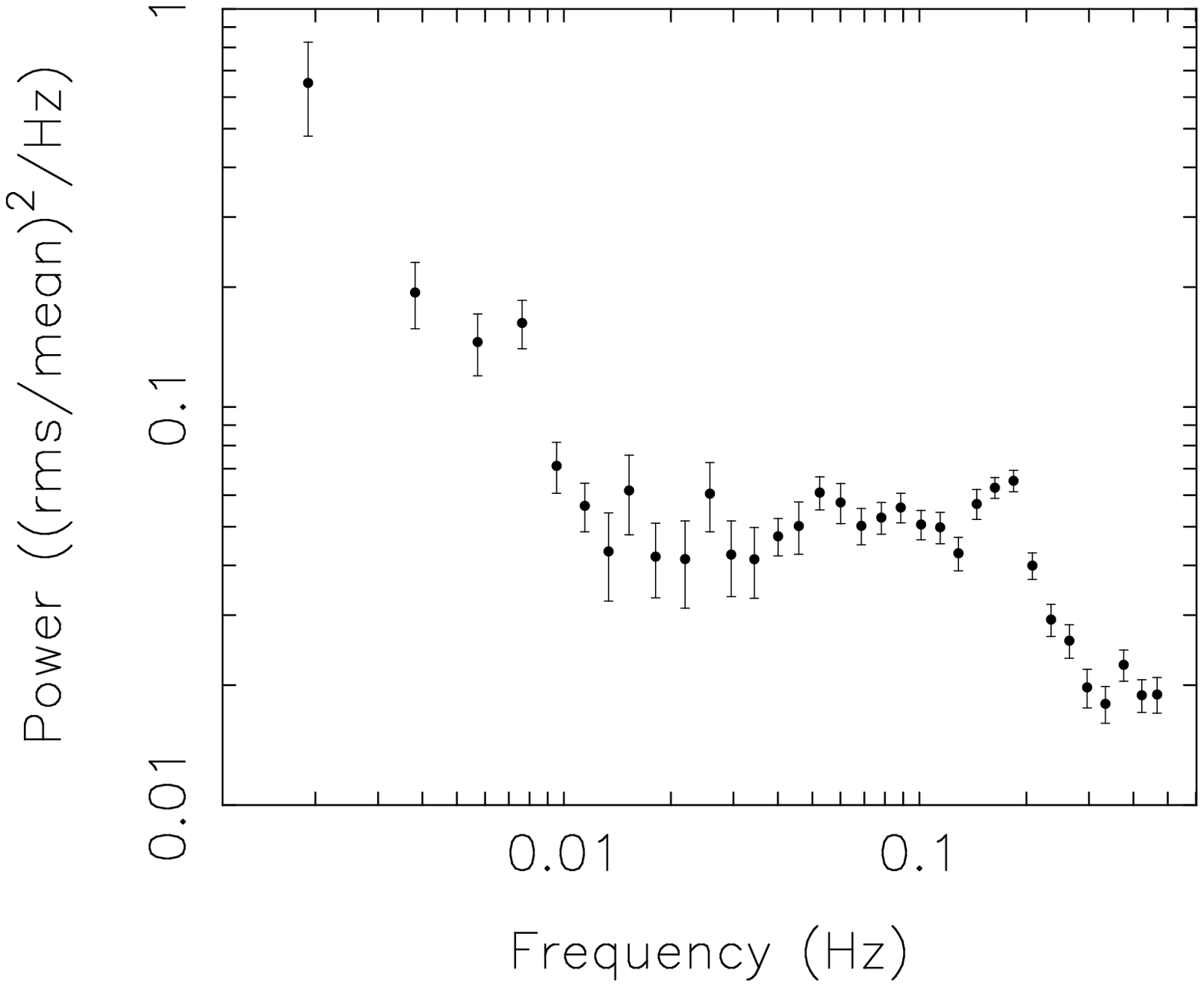,width=8.0cm,bbllx=179pt,bblly=60pt,bburx=565pt,bbury=529pt,angle=-90}
}
\caption[]{
The three-day averaged power spectrum (20--100 keV) for days 13--15 {\sl (left panel)} 
and the ten-day averaged power spectrum (20--100 keV) corresponding to days 51--60 of the X-ray 
outburst of \gro17\ {\sl (right panel)}. 
The frequency scale has been logarithmically rebinned into 34 bins. 
}
\label{pds_1719} 
\end{figure*}

The soft X-ray transient \gro17\ (=\,GRS~1716$-$249, Nova Oph 1993) was detected simultaneously 
with BATSE on board the {\sl Compton Gamma Ray Observatory}, and 
the SIGMA telescope on {\sl GRANAT}, on 1993 September 25 (Harmon \ea\ 1993a; Ballet \ea\ 1993). 
The source reached a maximum X-ray flux of $\sim$\,1.4 Crab (20--100 keV) within five days after first 
detection, and was remarkable for the stability of its hard X-ray emission on a time scale of days; 
its hard X-ray flux declined at a rate of $\sim$\,0.3\,$\pm$\,0.05\% per day (Harmon \ea\ 1993b). 
\gro17\ was detected above the BATSE 3$\sigma$ one-day detection threshold of 0.1 Crab (20--100 keV) 
for $\sim$\,80 days following the start of the X-ray outburst (Harmon \& Paciesas 1993). 
A time-series analysis of the hard X-ray variability of \gro17, observed with BATSE in the 20--100 keV energy 
band, was presented by van der Hooft \ea\ (1996). 
They analyzed the entire 80 day X-ray outburst of \gro17\ in the frequency interval 0.002--0.488 Hz. 
The power density spectra (PDSs) of \gro17\ show a significant peak, indicative of quasi periodic 
oscillations (QPOs) in the time series, whose centroid frequency increases from $\sim$\,0.04 Hz at 
the start of the outburst, to $\sim$\,0.3 Hz at the end. 
Van der Hooft \ea\ (1996) discovered that the evolution in time of the PDSs of GRO~J1719$-$24 can be 
described by a single characteristic profile. 
The evolution of the PDSs can be described as a gradual stretching by a factor $\sim$\,7.5 in frequency of 
the power spectrum, accompanied by a decrease of the power level by the same factor, such that the integrated 
power in a scaled frequency interval remains constant. 
Therefore, it is likely that the X-ray variability during the entire outburst of 
GRO~J1719$-$24 can be described by a single process, the characteristic time scale of which becomes shorter, but the 
fractional amplitude of which is invariant. 
This may be related to the strong anticorrelation of the break frequency and power density at the break observed 
in the PDSs of several black-hole candidates (Belloni \& Hasinger 1990). 
M\'{e}ndez \& van der Klis (1997) suggest a correlation with mass accretion rate may exist, i.e., the break 
frequency increases (and the power density decreases) with increasing mass accretion rate. 
Two average PDSs (20--100 keV) corresponding to days 13--15 and 51--60 of the X-ray outburst of 
\gro17\ are displayed in Figure~\ref{pds_1719}.

\gro17\ remained undetectable until 1994 September, when several X-ray flares were detected with both SIGMA and 
BATSE (Churazov \ea\ 1994; Harmon \ea\ 1994). 
Subsequent to strong X-ray flares in 1995 February (Borozdin, Alexandrovich \& Sunyaev 1995), a rapidly decaying 
radio flare was detected, followed by recurrent radio flaring activity (Hjellming \ea\ 1996). 
The relation between X-ray and radio events is similar to that observed in the superluminal radio-jet sources 
GRO~J1655$-$40 and GRS~1915$+$105 (Hjellming \ea\ 1996; Foster \ea\ 1996): radio emission follows the peak, or 
onset to decay of X-ray flares observed with BATSE in the 20--100 keV energy band, by intervals ranging from a few 
to 20 days (Hjellming \ea\ 1996). 
GRO~J1655$-$40 is a galactic black-hole candidate (BHC) with a dynamically determined mass of 
7.0\,$\pm$\,0.7 M$_{\odot}$ (Orosz \& Bailyn 1997; van der Hooft \ea\ 1998a).

A possible optical counterpart to the X-ray source was discovered by Della Valle, Mirabel \& Rodriquez (1994), 
the photometric and spectroscopic properties of which suggest that GRO J1719$-$24 is a low-mass X-ray binary. 
The optical brightness of \gro17, measured during three weeks after first X-ray detection, is modulated 
at a period of 0.6127 days, thought to be the superhump period (Masetti \ea\ 1996). 
Quiescent (optical) photometry and/or spectroscopy of \gro17\ has not been reported. 
The source is considered a black-hole candidate on the basis of its X-ray and radio similarities to dynamically 
proven BHCs. 

We have investigated the phase (or, equivalently, time) lags in the hard X-ray variability of \gro17\ during 
its 1993 X-ray outburst. 
We calculated lags between the 20--50 and 50--100 keV energy bands of the 1.024 sec time resolution BATSE data and 
compare our results with those obtained in recent similar studies of the black-hole candidates Cyg~X-1 
(Cui \ea\ 1997; Crary \ea\ 1998) and GRO~J0422$+$32 (Grove \ea\ 1997; van der Hooft \ea\ 1999).

\section{Analysis}

A time-series analysis of the hard X-ray (20--100 keV) data of the entire 1993 outburst of \gro17\ was 
presented by van der Hooft \ea\ (1996).
These data were obtained in two broad energy channels (20--50 and 50--100 keV) with the large-area detectors of 
BATSE, collected during 80 days following first X-ray detection on 1993 September 25. 
Fast Fourier Transforms were created for 524.288 sec long time intervals (512 time bins of 1.024 sec each); 
the corresponding frequency interval covered 0.002--0.488 Hz. 
The average number of uninterrupted 512 bin segments available with the source unocculted by the Earth was 
approximately 35 per day. 
See van der Hooft \ea\ (1996) for a detailed description of the reduction and analysis of these data. 

The complex Fourier cross spectra were created from the Fourier amplitudes in a way identical to that described by 
van der Hooft \ea\ (1999). 
These cross spectra were averaged daily.
Errors on the real and imaginary parts of the daily averaged cross spectra were calculated from the respective 
sample variances, and formally propagated when computing the phase and time lags. 
The phase lags, $\phi_{j}$, as a function of frequency were obtained from the cross spectra via 
$\phi_{j}$\,$=$\,arctan[Im($C_{j}^{12}$)/Re($C_{j}^{12}$)], and the corresponding time lag 
$\tau_{j}$\,$=$\,$\phi_{j}/2\pi\nu_{j}$, with $\nu_{j}$ the frequency in Hz of the $j$-th frequency bin. 
With these definitions, lags in the hard X-ray variations (50--100 keV) with respect to the soft X-ray variations 
(20--50 keV) appear as positive angles. 

\begin{figure*} 
\hbox{
\psfig{figure=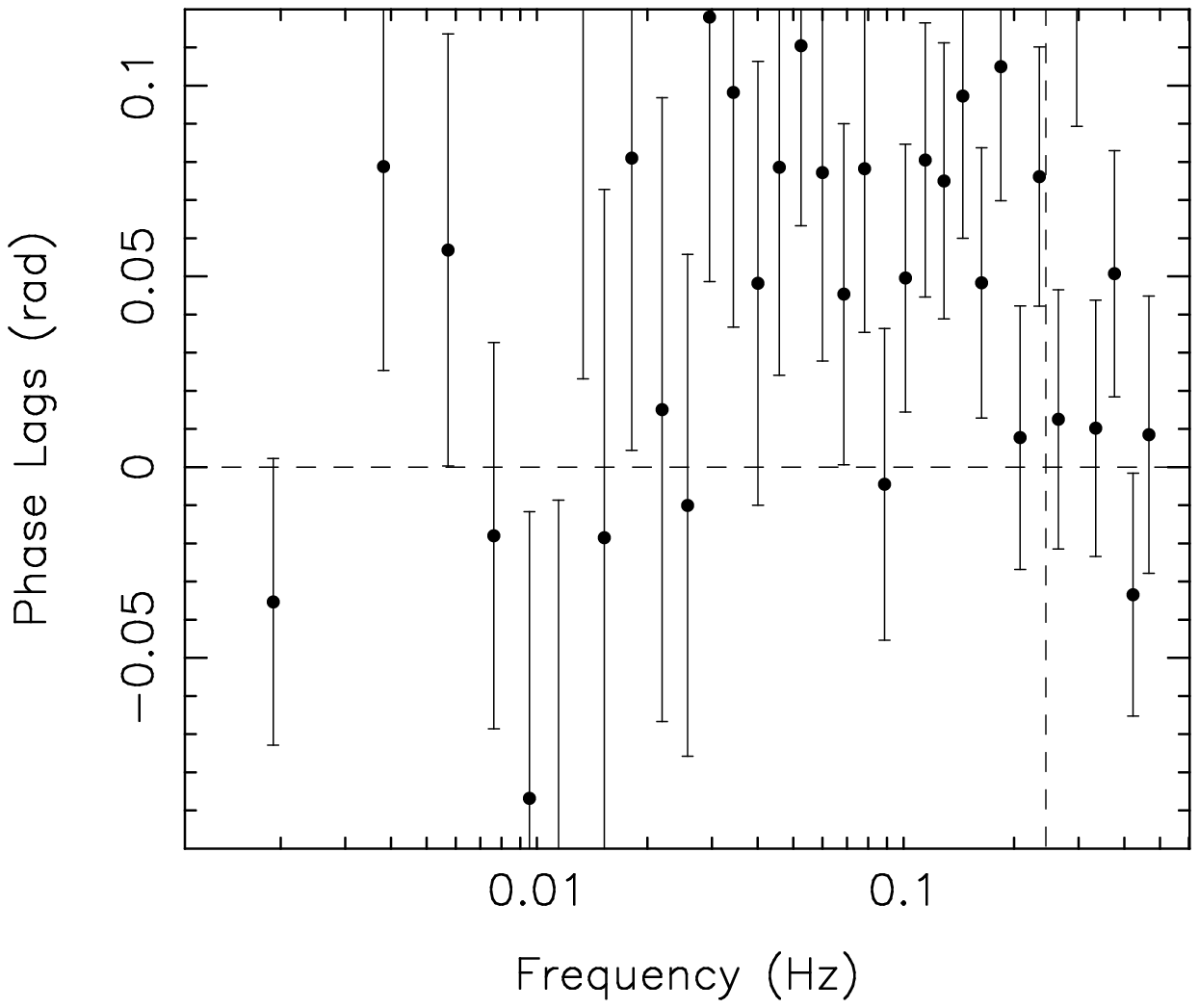,width=7.8cm,bbllx=240pt,bblly=95pt,bburx=537pt,bbury=449pt,angle=-90} 
\hspace{0.1cm}
\psfig{figure=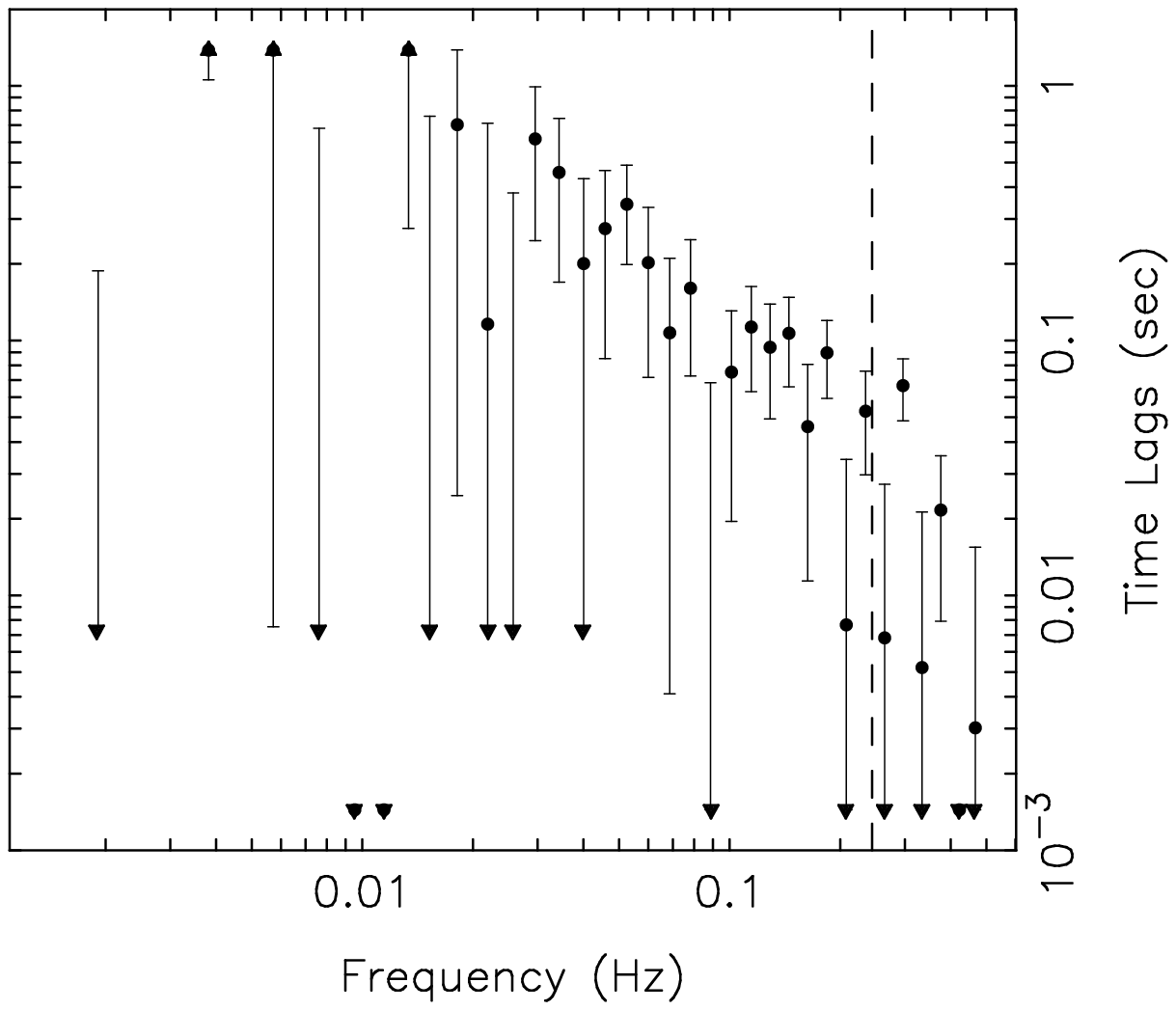,width=7.8cm,bbllx=240pt,bblly=143pt,bburx=537pt,bbury=497pt,angle=-90}
}
\caption[]{
Average phase {\sl (left panel)} and time {\sl (right panel)} lags of \gro17\ between the 20--50 and 50--100 keV 
energy bands (hard lags appear as positive angles). 
These averages cover the full 80 days outburst of the source. 
The frequency scale has been logarithmically rebinned into 34 bins.
Triangles indicate data points or error bars which are outside the scale of the figure. 
The vertical dashed lines indicate 0.5 $\nu_{\rm Nyq}$ above which the lags may be affected by data binning. 
}
\label{cross_1719} 
\end{figure*}

Cross spectra for a large number of days must be averaged and converted to lag values in order to obtain 
sufficiently small errors (see, e.g., Crary \ea\ 1998; van der Hooft \ea\ 1999). 
Therefore, we averaged the phase and time lags between the 20--50 and 50--100 keV energy bands of the entire 
80 day X-ray outburst of \gro17. 
These are presented in Figure~\ref{cross_1719}. 
The time lags are displayed on a logarithmic scale. 
Time lags at frequencies above 0.5 $\nu_{\rm Nyq}$ are displayed but not taken into account in our analysis, 
as Crary \ea\ (1998) have shown that data binning effects distort the shape of the cross spectra at 
these frequencies. 
These data show that at the lowest frequencies the phase lags are likely smaller than the high frequency lags 
(0.021\,$\pm$\,0.028 rad, average of 0.001--0.02 Hz; 9 bins). 
At frequencies above 0.02 Hz, the hard X-rays lag the soft by 0.072\,$\pm$\,0.010 rad (average of 0.02--0.20 Hz; 94 bins). 
The phase lags averaged over two 40 day intervals are similar to those averaged over the entire 80 day outburst, 
being 0.0017\,$\pm$\,0.028 rad and 0.041\,$\pm$\,0.043 rad, respectively, for the 0.001--0.02 Hz interval, 
and 0.082\,$\pm$\,0.013 rad and 0.061\,$\pm$\,0.016 rad, respectively, for the 0.02--0.20 Hz interval. 
The time lags of \gro17\ decrease with frequency as a power law, with index 1.04\,$\pm$\,0.13 for frequencies 
$\geq$\,0.01 Hz. 
The extrapolation of this power law for frequencies smaller than 0.01 Hz is well above the measured time lags. 

\section{Discussion}

The 20--100 keV energy spectrum steadily softened during the entire X-ray outburst of GRO J1719$-$24 in 1993; 
the photon index increased from 2.0 to 2.3\,$\pm$\,0.05 during the rise to peak intensity, beyond which 
the spectrum softened more gradually. 
No marked changes in the spectral shape were observed during the sudden decrease in X-ray flux in 1993 December 
(van der Hooft \ea\ 1996). 
It is not possible, on the basis of 20--100 keV BATSE observations alone to distinguish between black hole source 
states. 
However, observations at low X-ray energies during the decay of the X-ray light curve of \gro17, suggest that 
the source was most likely in the low (or hard) state. 
The 2--300 keV X-ray spectrum, obtained about 30 days after first detection of \gro17\ by combining SIGMA 
data with quasi-contemporaneous data taken by TTM on board {\sl Mir-Kvant}, was quite similar to the low state 
spectrum of Cyg~X-1.
The 2--300 keV spectrum of \gro17\ then had a power-law shape without a soft component, and a cut off at energies above 
100 keV (Revnivtsev \ea\ 1998). 
Therefore, these observations indicate that 30 days after the X-ray outburst had started, \gro17\ was in the low state. 
The lack of significant changes in the hard X-ray properties (van der Hooft \ea\ 1996) of \gro17, suggests that 
this conclusion applies to the entire 1993 outburst. 

Recently, Crary \ea\ (1998) and van der Hooft \ea\ (1999) have studied lags between the X-ray flux variations 
in 20--50 and 50--100 keV BATSE data of the black-hole candidates Cyg~X-1 and GRO~J0422$+$32. 
Cui \ea\ (1997) measured hard X-ray time lags in 2--60 keV {\sl RXTE} data of Cyg~X-1, obtained during 1996. 
Crary \ea\ (1998) studied Cyg~X-1 for a period of almost 2000 days, during which the source was likely in both the 
low, and high or intermediate state. 
They found that the lag spectra between the X-ray variations in the 20--50 and 50--100 keV energy 
bands of Cyg X-1 do not show an obvious trend with source state. 
They grouped the phase lag data according to the squared fractional rms amplitude 
of the noise, integrated in the frequency interval 0.03--0.488 Hz.  
They find that at the lowest frequencies the phase lag is consistent with zero. 
For higher frequencies the hard phase lag increases to a maximum of 0.04 rad near 0.20 Hz, and decreases again to near zero  
at the Nyquist frequency.

Crary \ea\ (1998) showed that binning effects decrease the observed hard X-ray time lags to zero 
at the Nyquist frequency. 
Therefore, time lags obtained for frequencies between 0.5 $\nu_{\rm Nyq}$ and $\nu_{\rm Nyq}$ may 
be affected by data binning. 
The Cyg~X-1 X-ray variations in the 50--100 keV band lag those in the 20--50 keV band over the 0.01--0.20 Hz 
frequency interval by a time interval proportional to $\nu^{-0.8}$. 

Cui \ea\ (1997) studied Cyg~X-1 during its 1996 spectral transitions. 
The observed period can be divided into a transition from the hard state to the soft state, a soft state, and a transition 
from the soft state back to the hard state. 
The lag spectra obtained by Cui \ea\ (1997) cover the frequency range 0.01--100 Hz. 
They find that during the state transitions the time lags between energy bands with average energy $E_0$ and $E_1$, scale 
with photon energy roughly as $\log$\,$(E_1/E_0)$. 
Such a scaling is consistent with the predictions of thermal Comptonization in the corona (see, e.g., Payne 1980; 
Hua \& Titarchuk 1996; Kazanas, Hua \& Titarchuk 1997). 
In the soft state the time lags become much smaller. 
This implies that in the soft state the size of the corona becomes much smaller. 

Van der Hooft \ea\ (1999) determined lags in the hard X-ray variability of GRO~J0422$+$32 during its 1992 outburst. 
Their time-series analysis covered the entire 180 day X-ray outburst. 
GRO~J0422$+$32 is a dynamically proven black-hole candidate; during its 1992 outburst it was most likely in the low 
state (van der Hooft \ea\ 1999). 
They averaged the phase lags of GRO~J0422$+$32 over a 30 day interval following first X-ray 
detection of the source, and over a flux-limited sample of the remaining data (95 days).
Statistically significant lags were derived for the shorter interval only.  
They find that at the lowest frequencies the phase lag of GRO~J0422$+$32 is consistent with zero 
(0.014$\pm$\,0.006 rad, 0.001--0.02 Hz).  
At frequencies $\geq$\,0.02 Hz, the variations in the 50--100 keV band lag those in the 20--50 keV band by 
0.039$\pm$\,0.003 rad (average of 0.02--0.20 Hz). 

The time lags of GRO~J0422$+$32, during the first 30 days of its outburst, decrease with frequency as a power law, 
with index $\sim$\,0.9 for $\nu$\,$>$\,0.01 Hz (van der Hooft \ea\ 1999). 
Grove \ea\ (1997) studied the time lags of GRO~J0422$+$32 between the X-ray variations in the 35--60 keV band and 
75--175 keV band with OSSE. 
They find that the hard X-ray emission lags the soft emission at all Fourier frequencies, decreasing roughly 
as $\nu^{-1}$ up to about 10 Hz. 
At frequencies of $\sim$\,0.01 Hz, hard time lags as large as 0.3 sec are observed. 
The hard time lags of GRO~J0422$+$32 obtained by Grove \ea\ (1997), are consistent with those obtained by van der 
Hooft \ea\ (1999).

The phase lags of \gro17\ are very similar to those of GRO~J0422$+$32 and Cyg~X-1. 
At frequencies below 0.02 Hz very small lags are observed (consistent with zero), while 
at frequencies of $\sim$\,0.10 Hz the variations in the 50--100 keV band lag those in the 20--50 keV band. 
However, the phase lags of \gro17, averaged in the interval 0.02--0.20 Hz, are about twice as large as those detected 
in GRO~J0422$+$32 and Cyg~X-1.

These results show that the hard time lags observed in \gro17, GRO~J0422$+$32 and Cyg~X-1 are all very similar. 
The hard X-ray radiation lags the soft by as much as $\sim$\,0.1--1 sec at low 
frequencies. 
The time lags are strongly dependent on the Fourier frequency, and decrease roughly as $\nu^{-1}$.
The $\nu^{-1}$ dependence of the hard time lags is very different from the lags expected from simple models of 
Compton upscattering of soft X-rays by a cloud of hot electrons near the black hole. 
In such a case, the energy of the escaping photons increases with the time they reside in the cloud. 
Therefore, higher energy photons lag the photons with lower energies by an amount proportional to the photon 
scattering time. 
If the hard X-rays are emitted from a compact region near the black hole, the resulting time lags should be 
independent of Fourier frequency and of the order of milliseconds.

Analysis of the hard time lags in the X-ray variability of black-hole candidates can provide information on the 
density structure of the accretion gas (Hua, Kazanas \& Titarchuk 1997).
Kazanas \ea\ (1997) argued that the Comptonization process takes place in an extended non-uniform 
cloud around the central source. 
They showed that such a model can account for the form of the observed PDS and energy spectra of compact sources. 
Hua \ea\ (1997) showed that the phase and time lags of the X-ray variability depend on the density 
profile of such an extended scattering atmosphere. 
Their Monte Carlo simulations of scattering in a cloud with a density profile proportional to $r^{-1}$ agree with 
our time lag data both in magnitude ($\sim$\,0.1 sec at 0.10 Hz) and frequency dependence ($\nu^{-1}$). 
The results presented here support the idea that the Comptonizing regions around the black holes in Cyg~X-1, 
GRO~J0422$+$32 and GRO~J1719$-$24 are quite similar in density distribution and size.

However, the observed lags require that the scattering medium has a size of order 10$^3$ to 10$^4$ 
Schwarzschild radii. 
It is unclear how a substantial fraction of the X-ray luminosity, which must originate from the conversion of 
gravitational potential energy into heat close to the black hole, can reside in a hot electron gas at such large 
distances. 
This is a generic problem for Comptonization models of the hard X-ray time lags. 
Also, such models do not specify the source of soft photons, nor do they account for the soft excesses and weak Fe 
lines seen in the energy spectra. 
Very detailed high signal-to-noise cross-spectral studies of the rapid X-ray variability of accreting BHCs, and 
combined spectro-temporal modeling may solve this problem.

\acknowledgments
FvdH acknowledges support by the 
Netherlands Foundation for Research in Astronomy with financial aid from the 
Netherlands Organisation for Scientific Research (NWO) under contract number 
782-376-011. 
FvdH also thanks the `Leids Kerkhoven--Bosscha Fonds' for a travel grant. 
CK acknowledges support from NASA grant NAG-2560. 
JvP acknowledges support from NASA grants NAG5-2755 and NAG5-3674. 
WHGL gratefully acknowledges support from the National Aeronautics and Space Administration. 
MvdK gratefully acknowledges the Visiting Miller Professor Program of 
the Miller Institute for Basic Research in Science (UCB). 
This project was supported in part by NWO under grant PGS 78-277.


\clearpage
%

%

\newpage

\begin{thebibliography}{} 

\bibitem[Ballet \ea\ 1993]{bal93} Ballet, J., Denis, M., Gilfanov, M., Sunyaev, R. 1993, IAU Circ., 5874

\bibitem[Belloni \& Hasinger 1990]{bell90} Belloni, T. \& Hasinger, G. 1990, A\&A, 227, L33

\bibitem[Borozdin, Alexandrovich \& Sunyaev 1995]{bor95} Borozdin, K., Alexandrovich, N., Sunyaev, R. 1995, IAUC., 6141

\bibitem[Churazov \ea\ 1994]{chu94} Churazov, E., Gilfanov, M.,  Ballet, J., Jourdain, E. 1994, IAU Circ., 6083

\bibitem[Crary \ea\ 1998]{crar98} Crary, D.J., Finger, M.H., Kouveliotou, C., van der Hooft, F., van der Klis, 
M., Lewin, W.H.G., van Paradijs, J. 1998, ApJ, 493, L71

\bibitem[Cui \ea\ 1997]{cui97} Cui, W., Zhang, S.N., Focke, W., Swank, J.H. 1997, ApJ, 484, 383

\bibitem[Della Valle, Mirabel \& Rodriquez 1994]{del94} Della Valle, M., Mirabel, I.F., Rodriquez, L.F. 
1994, A\&A, 290, 803

\bibitem[Foster \ea\ 1996]{fos96} Foster, R., Waltman, E.B., Tavani, M., Harmon, B.A., Zhang, S.N., Paciesas, W.S., 
Ghigo, F.D. 1996, ApJ, 467, L81

\bibitem[Grove \ea\ 1997]{gro97} Grove, J.E., Grindlay, J.E., Harmon, B.A., Hua, X.-M., Kazanas, D., McConnell, M. 1997, 
in AIP Conf. Proc. 410, ``Proceedings of the Fourth Compton Symposium'', eds. C.D. Derner, M.S. Strickman \& J.D. Kurfess 
(New York: AIP), 122

\bibitem[Harmon \ea\ 1993a]{har93a} Harmon, B.A., Zhang, S.N., Paciesas, W.S., Fishman, G.J. 1993a, IAU Circ., 5874

\bibitem[Harmon \ea\ 1993b]{har93b} Harmon, B.A., Fishman, G.J., Paciesas, W.S., Zhang, S.N. 1993b, IAU Circ., 5900

\bibitem[Harmon \& Paciesas 1993]{har93} Harmon, B.A. \& Paciesas, W.S. 1993, IAU Circ., 5913

\bibitem[Harmon \ea\ 1994]{har94} Harmon, B.A., Zhang, S.N., Paciesas, W.S., Wilson, C.A., Fishman, G.J. 1994, IAU Circ., 6104

\bibitem[Hjellming \ea\ 1996]{hjel96} Hjellming, H.M., Rupen, M.P., Shrader, C.R., Campbell-Wilson, D., 
Hunstead, R.W., McKay, D.J., 1996, ApJ, 470, L105

\bibitem[Hua \& Titarchuk 1996]{hua96} Hua, X.-M. \& Titarchuk, L. 1996, ApJ, 496, 280

\bibitem[Hua, Kazanas \& Titarchuk 1997]{hua97} Hua, X.-M., Kazanas, D., Titarchuk, L. 1997, ApJ, 482, L57

\bibitem[Kazanas, Hua \& Titarchuk 1997]{kaz97} Kazanas, D., Hua, X.-M., Titarchuk, L. 1997, ApJ, 480, 735

\bibitem[Masetti \ea\ 1996]{mas96} Masetti, N., Bianchini, A., Bonibaker, J., Della Valle, M., Vio, R. 
1996, A\&A, 314, 123

\bibitem[M\'{e}ndez \& van der Klis 1997]{mendez97} M\'{e}ndez, M. \& van der Klis, M. 1997, ApJ, 479, 926

\bibitem[Orosz \& Bailyn 1997]{oro97} Orosz, J.A. \& Bailyn, C.D. 1997, ApJ, 477, 876

\bibitem[Payne 1980]{pay80} Payne, D.G. 1980, ApJ, 237, 951

\bibitem[Revnivtsev, M. \ea\ 1998]{revn98} Revnivtsev, M., Gilfanov, M., Churazov, E., Sunyaev, R., 
Borozdin, K., Alexandrovich, N., Khavenson, N., Chulkov, I., Goldwurm, A., Ballet, J., Denis, M., 
Laurent, P., Roques, J.-P., Borrel, V., Bouchet, L., Jourdain, E. 1998, A\&A, 331, 557

\bibitem[van der Hooft \ea\ 1996]{fvdh96} van der Hooft, F., Kouveliotou, C., van Paradijs, J., Rubin, B.C., Crary, D.J., 
Finger, M.H., Harmon, B.A., van der Klis, M., Lewin, W.H.G., Norris, J.P., Fishman, G.J. 1996, ApJ, 458, L75

\bibitem[van der Hooft \ea\ 1998a]{fvdh98a} van der Hooft, F., Heemskerk, M.H.M, Alberts, F., van Paradijs, J. 
1998a, A\&A, 329, 538

\bibitem[van der Hooft \ea\ 1999]{fvdh99} van der Hooft, F., Kouveliotou, C., van Paradijs, J., Paciesas, W.S., 
Lewin, W.H.G., van der Klis, M., Crary, D.J., Finger, M.H., Harmon, B.A., Zhang, S.N. 1999, ApJ, 513, pp.

\end{thebibliography}
\end{document}